\documentclass[conference]{IEEEtran}
\usepackage{hyperref}
\usepackage{graphicx}%
\usepackage{multirow}%
\usepackage{amsmath,amssymb,amsfonts}%
\usepackage{amsthm}%
\usepackage{mathrsfs}%
\usepackage{xcolor}%
\usepackage{textcomp}%
\usepackage{manyfoot}%
\usepackage{booktabs}%
\usepackage{amssymb}
\usepackage{pifont}
\usepackage{listings}
\usepackage{listing}

\usepackage[ruled, lined, linesnumbered, commentsnumbered, longend]{algorithm2e}
\usepackage{enumitem}

\begin{document}
\title{Double Public Key Signing Function Oracle Attack on EdDSA Software Implementations}

\author{
    \IEEEauthorblockN{
        Sam Grierson\IEEEauthorrefmark{1},
        Konstantinos Chalkias\IEEEauthorrefmark{2},
        William J Buchanan\IEEEauthorrefmark{1},
        Leandros Maglaras\IEEEauthorrefmark{1}      
    }\\
    \IEEEauthorblockA{\IEEEauthorrefmark{1} 
        Blockpass ID Lab, Edinburgh Napier University, Edinburgh, UK\\
        Email: \{s.grierson2, b.buchanan, l.maglaras\}@napier.ac.uk
    }
    \IEEEauthorblockA{\IEEEauthorrefmark{2} 
        Mysten Labs Research
    }
}
\definecolor{BG}{gray}{0.95}

\maketitle
\begin{abstract}
EdDSA is a standardised elliptic curve digital signature scheme introduced to overcome some of the issues prevalent in the more established ECDSA standard. Due to the EdDSA standard specifying that the EdDSA signature be deterministic, if the signing function were to be used as a public key signing oracle for the attacker, the unforgeability notion of security of the scheme can be broken. This paper describes an attack against some of the most popular EdDSA implementations, which results in an adversary recovering the private key used during signing. With this recovered secret key, an adversary can sign arbitrary messages that would be seen as valid by the EdDSA verification function. A list of libraries with vulnerable APIs at the time of publication is provided. Furthermore, this paper provides two suggestions for securing EdDSA signing APIs against this vulnerability while it additionally discusses failed attempts to solve the issue.
\end{abstract}

\section{Introduction}
Since it was first proposed independently in the late '80s by Koblitz \cite{koblitz1987} and Miller \cite{miller1986}, Elliptic Curve Cryptography (ECC) has become the preferred choice for constructing classical public-key cryptosystems. The critical advantage of ECC is its capability to construct public key cryptosystems with a smaller key size than its discrete logarithm-based counterparts. For example, the Digital Signature Algorithm (DSA) proposed by the National Institute of Standards and Technology (NIST) for their Digital Signature Standard (DSS) (attributed to Kravitz \cite{kravitz1993}) has an ECC counterpart, the Elliptic Curve Digital Signature Algorithm (ECDSA) \cite{johnson2001}, which boasts greater efficiency and smaller key sizes while achieving similar levels of security. However, ECDSA is not without its share of common pitfalls that implementations can suffer from. For example, key recovery attacks are enabled by poorly generated random values \cite{nguyen2003} and nonce re-usage \cite{brengel2018}. Lattice-based attacks, such as those using the Lenstra-Lenstra-Lov\'asz (LLL) method \cite{lenstra1982}, have also been used to recover information about private keys from weak ECDSA signatures successfully \cite{poulakis2011, breitner2019}.

With the evident problems in ECDSA implementations and a loss of trust in NIST after the Snowden revelations, the cryptography community shifted towards a new cryptosystem based on Curve25519 proposed by Bernstein in 2006 \cite{bernstein2006}. In 2012, Bernstein \emph{et al.} \cite{bernstein2012} proposed using the Edwards variant of Curve25519 to construct a deterministic Schnorr-like \cite{schnorr1990} digital signature scheme. This scheme became the Edwards-curve Digitial Signature Algorithm (EdDSA). One of the main advantages of EdDSA over other ECC signature schemes is how the scalar multiplication of points on the curve can be implemented without branching and lookups depending on a secret value \cite{bernstein2012}. Due to its many advantages over ECDSA \cite{bernstein2015eddsa}, EdDSA quickly became widely implemented and was eventually standardised in both RFC 8032 \cite{josefsson2017} and NIST's own FIPS 186-5 \cite{moody2023}.

This paper discloses an undiscovered vulnerability related to the implementation of EdDSA. This vulnerability is severe enough that adversaries can easily exploit it to extract the private key during the EdDSA signing process. This attack requires that an adversary use the signing function as an oracle that expects arbitrary public keys as inputs. While the majority of applications that use EdDSA are unlikely to expose signing functions to end users publicly or may mitigate the issue before signing invocation, there are some applications in which private and public keys are managed in different ways, exposing the surface to attack by adversaries. The details of this attack are given later in the paper, along with ways to mitigate the attack easily.

The rest of this paper is organised as follows: In the remainder of this section, various work related to EdDSA and its vulnerabilities is outlined, and the contributions of this paper are specified. In Section \ref{sec:eddsa} provides background information on the EdDSA algorithm required for the rest of the paper. Section \ref{sec:attack} describes the double public key signing function oracle attack and gives a list of libraries with EdDSA implementations vulnerable to the attack. In Section \ref{sec:countermeasures}, possible countermeasures against the described vulnerability are given and Section \ref{sec:conclusion} concludes the paper.

\subsection{Related Work}

Since the proposal of Ed25519 by Bernstein \emph{et al.} \cite{bernstein2012} in 2012 and the subsequent generalisation of the algorithm into EdDSA \cite{bernstein2015eddsa}, there has been a significant amount of work detailing various formal security notions of EdDSA as well as attacks to both the algorithm itself and implementation of the algorithm. Due to the fact that its construction is heavily based on the Schnorr signature scheme \cite{schnorr1990}, security of the schemes proposed by Bernstein \emph{et al.} in \cite{bernstein2012} and \cite{bernstein2015} is based on similar assumptions. More recently, Brendel \emph{et al.} \cite{brendel2020} gave a comprehensive security analysis of Ed25519 based on its implementation as per the RFC 8032 standard \cite{josefsson2017}, in which they found that certain implementations guarantee stronger security than others. Furthermore, work by Chalkias \emph{et al.} \cite{chalkias2020} was done to formalise EdDSA implementations under the strictest notions of security.

There have also been some high-profile attacks against EdDSA. In 2017, an issue arose in an implementation of Ed25519 used by the Monero crypto-currency, which allowed users to get around double-spending prevent ions. This issue was mitigated by checking the order of the key using full scalar multiplication and arose due to the unique way in which Monero used Ed25519 \cite{luigi2017}. Samwel \emph{et al.} \cite{samwel2018} demonstrated that differential power analysis could be used on Ed25519's underlying hash function SHA-512. In particular, their work targeted the WolfSSL implementation and required 4000 EM traces to be successful. In an extension to this work, Weissbart \emph{et al.} \cite{weissbart2019} used machine learning techniques to reduce this attack to a single EM trace.

Another type of attack against EdDSA is a fault attack. Romallier and Pelissier \cite{romailler2017} demonstrated that a single fault in the EdDSA signing process could be used to recover enough private key material for an attacker to sign arbitrary messages. Poddebniak \emph{et al.} \cite{poddebniak2018} also studied fault attacks against deterministic digital signature schemes such as EdDSA, formalising requirements for protocols to be vulnerable to these types of attacks. Approaching the same problem slightly differently, Cao \emph{et al.} \cite{cao2022} constructed lattice-based attacks to recover private key information from deterministic digital signature schemes vulnerable to fault attacks.

This work presents an attack on the standard rather than an attack on implementation-specific details found in EdDSA software or hardware. More specifically, the standards fail to specify the format of key input into the EdDSA signing function. Due to the algorithmic details, if an adversary was able to use the signing function as an Oracle expecting arbitrary public key inputs, then it is possible for them to recover the full private key trivially. To the best of the authors' knowledge, this issue was unreported until now.

\subsection{Contributions}

The main contributions of this paper are as follows:
\begin{itemize}
    \item A new attack against the EdDSA standards RFC 8032 \cite{josefsson2017} and FIPS 186-5 DSS \cite{moody2023} is presented. It is shown that unless necessary precautions are taken, an adversary can perform full private key recovery if given Oracle access to the EdDSA signing function
    \item A list of potentially unsafe EdDSA libraries is given. At the time of writing, there are 45 libraries impacted by this vulnerability. Misuse of these libraries can result in private key exposure. Currently, 8 of the 45 impacted libraries have implemented fixes to the issues after notification.  
    \item Finally, two countermeasures against this type of attack are given. These countermeasures are simple changes to the vulnerable EdDSA software implementations found in many libraries. Both changes require only a small amount of additional overhead in the signing function. 
\end{itemize}

\section{Edwards-curve Digital Signature Algorithm}
\label{sec:eddsa}

EdDSA is a digital signature algorithm similar to ECDSA \cite{johnson2001} proposed by Bernstein \emph{et al.} \cite{bernstein2012}. RFC 8032 \cite{josefsson2017} defines EdDSA defines for two twisted Edwards curves Ed25519 (similar to curve25519 proposed by Bernstein \cite{bernstein2006}) and Ed448; however, EdDSA may be instantiated over other curves. For a fixed field $k$, a twisted Edwards curve with the coefficients $a, d \in k$ is the curve:
\[
    E_{a,d} : ax^2 + y^2 = 1 + dx^2y^2
\]
where $a$ and $b$ are non-zero elements. For example, Ed25519 (curve25519) is defined over $\mathbb{F}_p$ where $p = 2^{255} - 19$. Similarly, Ed448 is defined over $\mathbb{F}_p$ where $p = 2^{448} - 2^{244} - 1$ and offers a 224-bit security level compared to the 128-bit security of Ed25519.

The sum of two points in the Ed25519 and Ed448 curves is represented by the following addition rule:
\[
    (x_1, y_1) + (x_2, y_2) = \left(\frac{x_1 y_2 + y_1 x_2}{1 + d x_1 x_2 y_1 y_2}, \frac{y_1 y_2 - a x_1 x_2}{1 - d x_1 x_2 y_1 y_2}\right)
\]
where the point $(0, 1)$ is the neutral element. The above rule can also be used for point doubling and addition. Adding a point $P$ to itself $n$ times is the same as multiplication by a scalar, denoted as $n \cdot P$.

\begin{table}[ht]
\centering
\caption{Parameters for Ed25519 and Ed448 (RFC 8032 \cite{josefsson2017})}
\label{tab:params}
\begin{tabular}{l l l}
\toprule
\textbf{Parameter} & \textbf{Ed25519} & \textbf{Ed448} \\
\midrule
Field Modulus ($p$) & $2^{255} - 19$ & $2^{448} - 2^{244} - 1$ \\
Key Bits ($b$) & $256$ & $456$ \\
Hash Function ($\text{H}$) & SHA-512 & SHAKE256 \\
Cofactor ($c$) & 8 & 4 \\
Coefficient ($d$) & $-121665/121666$ & $-39081$ \\
Coefficient ($a$) & $-1$ & $1$ \\
Base Point ($G$) & $(x, y) \in \mathbb{F}^2_p$ (see \cite{langley2016}) & $(x, y) \in \mathbb{F}^2_p$ (see \cite{langley2016}) \\
Curve Order ($\ell$) & see \cite{langley2016} & see \cite{langley2016} \\
\bottomrule
\end{tabular}
\end{table}

EdDSA uses a Fiat-Shamir transformed Schnorr-like identification protocol to generate a cryptographic signature based on elliptic curve point addition. The EdDSA protocol is standardised under the Ed25519 and Ed488 curves; the parameters for both curves are given in Table \ref{tab:params}. Note that the actual number of points on the curve is $\lvert E_{a,d} \rvert = c \cdot \ell$. The presence of the cofactor $c$ in the order of the curve makes it harder to use in applications where prime-order groups are required for  cryptographic proof.

The elliptic curve group $E_{a,d}$ is isomorphic to $\mathbb{Z}_\ell \times \mathbb{Z}_c$, where a base point $G \in E_{a,d}$ generates a subgroup of order $\ell$ and a small torsion point $T_c \in E_{a,d}$ generates the subgroup of order $c$. Any point $P \in E_{a,d}$ can be uniquely represented as the linear combination of $G$ and $T_c$ with $P = g \cdot B + t \cdot T_c$ where $0 \leq g < \ell$ and $0 \leq t < c$. In this case, the discrete log of $P$ base $G$ is $g$. $P$ is small order if $g = 0$, mixed order if $t \neq 0$ and $g \neq 0$, and order $\ell$ if $g \neq 0$ and $t = 0$.

\subsection{EdDSA Signing}

As defined in the RFC 8032 standard \cite{josefsson2017}, EdDSA uses a $b$-bit long private key $pk$ and a hash function $\text{H}$ that produces a $2b$-bit output. An integer $s$ is generated by taking the hash of the secret key $\text{H}(sk) = (h_0, h_1, \ldots, h_{2b - 1})$ and computing $s = 2^{b - 1} + \sum_{3 \leq i \leq b - 3} 2^i h_i$. The public key $pk$ is then computed by taking the curve's base point $G$ as defined in the public parameters and computing $pk = s \cdot G$.

\begin{algorithm}[ht]
\caption{EdDSA Signing}
\label{alg:sign}
\KwIn{$m$, $\text{H}$, $sk$, $G$, and $pk$}
\KwOut{The signature $(R, S)$}
\vspace{1.5mm}
$h := \text{H}(sk)$ \\
$s := 2^{b - 1} + \sum_{3 \leq i \leq b - 3} 2^i h_i$ \\
$r := \text{H}(h_b, \ldots, h_{2b - 1} \mid \mid m) \pmod{\ell}$ \\
$R := r \cdot G$ \\
$S := r + \text{H}(R \mid \mid pk \mid \mid m) \cdot s \pmod{\ell}$ \\
\textbf{return} $(R, S)$
\end{algorithm}

The signature $(R, S)$ of a message $m \in \{0, 1\}^\ast$ is computed according to Algorithm \ref{alg:sign}. A significant difference between EdDSA and ECDSA is the signature generated is deterministic in EdDSA; in other words, for a message, any signature computed using the same key pair and public parameters will always be the same.

Both signatures and keys can be encoded for space-efficient transmission. According to the RFC 8032 standard \cite{josefsson2017}, an element of the scalar field $\pmod{\ell}$ is encoded with a 256-bit little-endian string. If the scalar is reduced $\bmod\ \ell$ it is considered to be a canonical encoding; otherwise, it is non-canonical. A point $P = (x, y) \in E_{a, d}$ is also encoded as a 256-bit string, with 255 bits devoted to the encoding of $y$ in little-endian format and a single bit devoted to the encoding the sign of $x$. Given a serialisation of $P$ the $x$ coordinate is restored \emph{via} $x := \pm \sqrt{(y^2 - 1)/(dy^2 + 1)}$. If the $y$ coordinate is reduced mod $p$ encoding is canonical; otherwise, it is non-canonical.

\subsection{EdDSA Signature Verification}

The EdDSA verification algorithm given in Algorithm \ref{alg:vrfy} generally conforms to both the RFC 8032 standard and \cite{josefsson2017}, and the NIST FIPS 186-5 standard \cite{moody2023}, while also providing the strongest notion of security defined by Brendel \emph{et al.} \cite{brendel2020} and Chalkias \emph{et al.} \cite{chalkias2020}: Strong UnForgeability under Chosen Message Attacks (SUF-CMA) with Strongly Binding Signatures (SBS). This means that efficient adversaries cannot output valid signatures on new messages nor find a new signature for old messages. Furthermore, messages are bound to the public key, a property shown to be lacking in the RFC 8032 variant of EdDSA \cite{brendel2020}.

The verification algorithm given in Algorithm \ref{alg:vrfy} performs several checks to ensure the scheme's security against various attacks. First, the generic check to ensure that the public key $pk$ and the point $R$ from the signature $(R, S)$ are valid points on the curve $E_{a,d}$ is performed. The algorithm then ensures that the scalar value $S$ is not any of the values $0, \ldots, \ell - 1$, since $S' := S + n \cdot \ell$ would also satisfy the verification algorithm for $n \in \mathbb{Z}$, Checking the value of $S$ ensures that the scheme satisfies the requirements for SUF-CMA \cite{brendel2020}. The algorithm then rejects any non-canonical encodings of $pk$ and $R$. Rejecting non-canonical encodings is required by both RFC 8032 \cite{josefsson2017} and FIPS 186-5 \cite{moody2023}.

\begin{algorithm}[ht]
\caption{EdDSA Verification}
\label{alg:vrfy}
\KwIn{$m$, $(R,S)$, $\text{H}$, $G$, and $pk$}
\KwOut{$b \in \{0, 1\}$}
\vspace{1.5mm}
\If{$pk \notin E_{a,d}$ \textbf{\emph{or}} $R \notin E_{a,d}$}{\textbf{return} $0$}
\If{$S \notin \{0,\ldots \ell - 1\}$ \textbf{\emph{or}} $\lvert pk \lvert \geq \ell$ \textbf{\emph{or}} $\lvert R \lvert \geq \ell$}{\textbf{return} $0$}
\If{$pk = t \cdot T_c$ \emph{for} $\forall t \in \{0, \ldots, c - 1\}$ \emph{and} $\forall T_c \in E_{a, d}$}{\textbf{return} $0$}
\If{$c \cdot S \cdot G = c \cdot R + c \cdot \text{\emph{H}} (R \mid \mid pk \mid \mid m) \cdot pk$} {\textbf{return} $1$}
\textbf{return} $0$
\end{algorithm}

The final check ensures that the public key $pk$ used to sign the message is not one of a set of small order points on the curve $E_{a, d}$. This check is not part of any standard and rarely appears in practical implementations. This additional check aims to ensure public keys are strongly bound to the signature to achieve the SBS security notion. This is because if $pk$ is a $c$-torsion point, an adversary can any value for their signature such that $S \cdot G = R$ and the resulting signature verifies under any message. Bernstein \emph{et al.} \cite{bernstein2012} identified this vulnerability in work but regarded it as unproblematic. However, some specific cases have arisen where checking for small-order keys becomes important, specifically for building specialised protocols \cite{brendel2020}. While this may seem a cumbersome addition to the verification process, the number of small-order public keys is quite small and can be pre-computed and stored for fast verification \cite{chalkias2020}.

Verification of an EdDSA signature can be done either cofactored or cofactorless. The verification described by Algorithm \ref{alg:vrfy} is cofactored. If the implementation uses cofactorless implementation, then it is required to reduce $\text{H}(R \mid \mid pk \mid \mid m)$ to the range $[0, \ell)$ before multiplication by $pk$. Not doing so may cause implementations to disagree on the validity of signatures generated by mixed-order public keys. When performing cofactored verification, multiplication by $c$ should be performed as a separate scalar-by-point multiplication. Failing to ensure separate scalar-by-point multiplication can cause the result in $c \cdot \text{H}(R \mid\mid pk \mid\mid m) \bmod \ell$ not being divisible by $c$ and thus, not clear the low order component in $pk$ if it exists. While Bernstein \emph{et al.} \cite{bernstein2012} originally proposed to use cofactorless verification, EdDSA standards recommend using the cofactored verification algorithm \cite{josefsson2017, moody2023}.

\section{Double Public Key Signing Function Oracle Attack}
\label{sec:attack}

The discovered vulnerability takes the form of an oracle attack. The oracle uses the signing function of a deterministic signature scheme with a fixed secret key and message parameters to compute a signature given an arbitrary public key. If given access to this type of oracle, an adversary can use it to recover the secret key by submitting two different public keys. In this section, an attack methodology is described, and a list of affected libraries and their current status on fixing the issue is given.

\subsection{Attacking EdDSA}

According to both the RFC 8032 \cite{josefsson2017} and FIPS 186-5 \cite{moody2023} standards, EdDSA signatures are deterministic. This means that for the same message $m \in \{0, 1\}^\ast$ input to a signing function with public key $pk$ and secret key $sk$, a unique signature $(R, S)$ is generated. An important detail of the signing function given in Algorithm \ref{alg:sign} is that the signer's public key is used in the deterministic computation of the scalar value $S$ but not the point on the curve $R$ in the signature $(R, S)$. The implication of this is that if an adversary was able to use the signing function as an oracle that expects arbitrary public key inputs, they could compute two signatures $(R, S)$ and $(R, S')$ corresponding to the same $m$.

Assuming access to a signing oracle $\mathcal{O}_{\textsf{sign}_{sk, m}}$ with fixed parameters $m$ and $sk$, the adversary would perform the following steps to recover $sk$:
\begin{description}[leftmargin=0cm]
    \item[Step 1:] The adversary queries the oracle with two public keys $pk$ and $pk'$. The public keys need not be paired to the fixed secret key $sk$. The resulting signatures share the same $R$ value and differ on the $S$ values.
    \begin{description}[font=\normalfont, labelwidth=0.6cm, leftmargin=0.8cm]
        \item[1.1] Compute $pk := s \cdot G$ and $pk' := s' \cdot G$ where $s \neq s'$. Note that $pk$ and $pk'$ should satisfy the requirements of the EdDSA scheme.
        \item[1.2] Query $\mathcal{O}_{\textsf{sign}_{sk, m}}$ with $pk$ and $pk'$ and recieve the two signatures $\sigma$ and $\sigma'$.
        \item[1.3] Check that for $\sigma = (R, S)$ and $\sigma' = (R', S')$ the values $R = R'$. 
    \end{description}

    \item[Step 2:] With the two signatures $\sigma$ and $\sigma'$ corresponding to $pk$ and $pk'$ the adversary can now attempt to recover $sk$. When signing a message, the signing algorithm computes the $S$ value as  $S = r + \text{H}(R \mid\mid pk \mid\mid m) \cdot s \pmod \ell$. Because $r$ is derived from $sk$ which is the same for both signatures, and $R$, $pk$, and $m$ are all known to the adversary, they can use this to compute $s$.
    \begin{description}[font=\normalfont, labelwidth=0.6cm, leftmargin=0.8cm]
        \item[2.1] The adversary computes $e := \text{H}(R \mid\mid pk \mid\mid m)$ and $e' := \text{H}(R \mid\mid pk' \mid\mid m)$.
        \item[2.2] Because $S = r + e \cdot s \pmod \ell$ and $S' = r + e' \cdot  s \pmod \ell$, subtracting $S'$ from $S$ gives the following
        \[
            \begin{aligned}
                S - S' &= r + e \cdot s - r + e' \cdot s &\pmod \ell \\
                       &= e \cdot s - e' \cdot s         &\pmod \ell \\
                       &= s \cdot (e - e')               &\pmod \ell
            \end{aligned}
        \]
        \item[2.3] Dividing $S - S'$ through by $e - e'$ to recovers the value
        \[
            s = (S - S')(e - e')^{-1} \pmod \ell
        \]
    \end{description}
\end{description}

After completing step 2, the adversary has access to the secret integer $s$, which can be used to arbitrarily compute values of $S$. Even if $s$ is known, it remains impossible to compute a $r$ value for a new message since the values $h_b,\ldots,h_{2b-1}$ are unknown to the adversary. However, selecting any random value of $r$ can computing a new signature $\sigma = (R, S)$ for any message $m$ still satisfies
\[
    \begin{aligned}
        c \cdot S \cdot G &= c \cdot (r + \text{H}(R \mid\mid pk \mid\mid m) \cdot s) \cdot G \\
                          &= c \cdot R + c \cdot \text{H}(R \mid\mid pk \mid\mid m) \cdot s \cdot G \\
                          &= c \cdot R + c \cdot \text{H}(R \mid\mid pk \mid\mid m) \cdot pk
    \end{aligned}
\]
meaning the verification algorithm still holds. Therefore, it is still possible to sign arbitrary messages, effectively breaking the SUF-CMA security with the SBS notion of security that EdDSA guarantees.

\subsection{Vulnerable Libraries}

There are a huge number of software implementations of EdDSA across many different languages. To give an idea of how common this vulnerability can be, a table of vulnerable libraries can be seen in Table \ref{tab:libs}. Most of these libraries are taken from the IANIX list of ``Things that use Ed25519'' \cite{ianix2023}. These libraries have been notified of the issues, and their current status of fixing the vulnerability at the time of publication is included in the table \footnote{A comprehensive list of libraries and requested fixes can be found here: \url{https://github.com/MystenLabs/ed25519-unsafe-libs}}.

\begin{table}[ht]
\centering
\caption[Affected libraries]{Libraries affected by the double public key signing function oracle attack (at time of evaluation).}
\label{tab:libs}
\begin{tabular}{l l c}
\toprule
\textbf{Library} & \textbf{Language} & \textbf{Fixed} \\
\midrule
OpenGNB & C & \ding{55} \\
GNU Nettle & C & \ding{55} \\
iroha-ed25519 (Hyperledger Project) & ASM/C & \ding{55} \\
ed25519-donna (Andrew Moon) & C & \ding{55} \\
ed25519 (Orson Peters) & C & \ding{55} \\
libbrine (Kevin Smith) & C & \ding{55} \\
Ed25519 (ArduinoLibs) & C++ & \ding{55} \\
Trezor firmware & C & \ding{51} \\
Harbour (Viktor Szakats) & C & \ding{51} \\
ed25519 (Hans Wolff) & C\# & \ding{55} \\
Ed25519 (CryptoManiac) & C\# & \ding{55} \\
ed25519-dalek & Rust & \ding{51} \\
polkadot-js/wasm & Rust/WASM & \ding{51} \\
ed25519\_dart (Oleksii Semeshchuk) & Dart & \ding{55} \\
riclava\_ed25519 (riclava) & Dart & \ding{55} \\
ed25519 (Kevin Downey) & Clojure & \ding{55} \\
hs-scraps (Vincent Hanquez) & Haskell & \ding{55} \\
ed25519-java (k3d3) & Java & \ding{55} \\
ed25519 (Bjorn Arnelid) & Java & \ding{55} \\
Punisher.NaCl (Arpan Jati) & Java & \ding{55} \\
ED25519 (Mick Michalski) & Java & \ding{55} \\
vRallev/ECC-25519 & Java & \ding{55} \\
ed25519-elisabeth & Java & \ding{51} \\
Crypt::Ed25519 (Marc Lehmann) & Perl & \ding{55} \\
ed25519.py (Ed25519 authors) & Python & \ding{55} \\
pyca/ed25519 & Python & \ding{55} \\
python-pure25519 (Brian Warner) & Python & \ding{55} \\
nmed25519 (naturalmessage) & Python & \ding{55} \\
ed25519.py (Shiho Midorikawa) & Python & \ding{55} \\
py-ed25519-bindings & Python & \ding{55} \\
ed25519swift (pebble8888) & Swift & \ding{55} \\
supercop.js (1p6 Flynx) & JS & \ding{55} \\
substack/ed25519-supercop & JS & \ding{55} \\
libeddsa (Philipp Lay) & C & \ding{55} \\
SommerEngineering/Ed25519 & C\# & \ding{55} \\
ChorusOne/solanity & CUDA & \ding{55} \\
ncme/c25519 & C & \ding{55} \\
luazen (Phil Leblanc) & C & \ding{55} \\
amber (Pelayo Bernedo) & C++ & \ding{55} \\
FLD ECC AVX2 & C & \ding{55} \\
mwmiller/ed25519\_ex & Elixir & \ding{55} \\
php-ed25519-ext & PHP & \ding{55} \\
niv/ed25519.nim (Bernhard Stöckner) & Nim & \ding{55} \\
mipher (Marco Paland) & Typescript & \ding{55} \\
Monocypher & C & \ding{51} \\
LuaMonocypher & Lua & \ding{55} \\
monocypher.cr & Crystal & \ding{55} \\
py\_ssh\_keygen\_ed25519 (Péter Szabó) & Python & \ding{55} \\
KinomaJS & Javascript & \ding{55} \\
erlang-libdecaf & Erlang & \ding{51} \\
gen-ed25-keypair & Haskell & \ding{55} \\
horse25519 (Yawning Angel) & C & \ding{51} \\
\bottomrule
\end{tabular}
\end{table}

\section{Countermeasures}
\label{sec:countermeasures}

Fortunately, due to the nature of the oracle attack, the majority of applications with dependencies on the libraries listed in Table \ref{tab:libs} probably are safe due to not publicly exposing affected signing functions. That said, due to the nature of these libraries, a user can inadvertently expose the attack surface when building their application, as was the case with Monero in 2017 \cite{luigi2017}. Therefore, it is recommended that the implementer of the EdDSA standard follow one of two methods to prevent this attack: Correctly storing the public key along with the secret key or re-deriving the public key from the secret key each time the signing function is invoked.

\begin{listing}
\begin{lstlisting}[label=lst:opengnb, language=c,caption='OpenGNB public-key signature C interface']

void ed25519_sign(unsigned char* signature,
                  unsigned char const* message, 
                  size_t message_len,
                  unsigned char const* public_key,
                  unsigned char const* private_key);
 \end{lstlisting}
\caption{OpenGNB public-key signature C interface.}
\label{lst:opengnb}
\end{listing}

The double public key signing function oracle attack occurs primarily due to insecure APIs around the signing function of the deterministic signature scheme. For example, the Ed25519 signing function taken from the OpenGNB library shown in Listing \ref{lst:opengnb} has two separate arguments for the public and private keys. If an application using this library exposes this API publicly or mishandles the management of the keys, it could expose itself to the attack. The solution to this is to redesign the API to ensure that the secret and public keys pair are always tied together. It is common practice in many libraries to only accept a secret key in the signing function. For example, the Ed25519 signing function taken from the Libsodium library shown in Listing \ref{lst:libsodium} accepts only the signature, message, and secret key as an argument. However, the public key is still required by the signing algorithm. There are two solutions to this.

\begin{listing}[ht]
\begin{lstlisting}[label=lst:libsodium, language=c,caption='Libsodium public-key signature C interface']

int crypto_sign(unsigned char* sm,
                unsigned long long* smlen_p,
                unsigned char const* m, 
                unsigned long long mlen, 
                unsigned char const* sk);           
\end{lstlisting} 
\caption{Libsodium public-key signature C interface.}
\label{lst:libsodium}
\end{listing}

\subsection{Correct Key Storage}
\label{subsec:store}

The simplest solution is to ensure the public and private keys are stored together and accept them as a single argument to the signing function. This is also slightly more efficient computationally than the other option. Both the public and private keys for EdDSA are 32 or 57-byte for Ed25519 and Ed448, respectively. The solution found in the majority of libraries is to generate the public-private keypair and store the secret key as a 64-byte string. The first 32 bytes are the private key, with the remaining 32 bytes being the public key. The main downside of this is the private key is now 64 or 114 bytes for Ed25519 and Ed448. However, this increased storage space should be acceptable in all but the most extreme cases.

\begin{listing}[ht]
\begin{lstlisting}[label=lst:safekeygen, language=c,caption='An example of safe Ed25519 key generation and storage']

void ed25519_keypair(unsigned char* pk,
                     unsigned char* sk) {
    unsigned char seed[32];
    randombytes(seed, 32);
    sha512(sk, seed, 32);
    gen_pk(pk, sk);
    memmove(sk, seed, 32);
    memmove(sk + 32, pk, 32);
}

\end{lstlisting}
\caption{An example of safe Ed25519 key generation and storage.}
\label{lst:safekeygen}
\end{listing}

In Listing \ref{lst:safekeygen}, an example of safe key generation and storage is given based on the Libsodium Ed25519 implementation. In this example, the $seed$ byte array is stored as the first 32 bytes of the private key $sk$. This means that the users of the library can retrieve the initial random seed used to generate the public and private key pair. When invoking the signing function, the secret key would be derived by taking the first 32 bytes and hashing them with the SHA-512 hash function, and the public key would be the remaining 32 bytes.

\subsection{Public Key Re-derivation}

The other option is to receive the public key on every invoke of the signing function. Obviously, this consumes significantly more CPU cycles in the long term than storing the public key alongside the private key, as suggested in Section \ref{subsec:store}. However, the additional space requirements are no longer necessary, which may be more suitable for use cases with extreme memory restrictions. This solution is far less common to see in software implementations of EdDSA.

\begin{listing}[ht]
\begin{lstlisting}

void ed25519_sign(unsigned char* sig,
                  unsigned char const* m,
                  size_t mlen,
                  unsigned char const* sk) {
    unsigned char s[32];
    unsigned char pk[32];
    sha512(s, sk, 32);
    gen_pk(pk, s);
    ...
}

\end{lstlisting}
\caption{An example of signing with public key re-derivation.}
\label{lst:safesign}
\end{listing}

In Listing \ref{lst:safesign}, an example of an Ed25519 signature function with key re-derivation is given. In this example, the secret key $sk$ given is expected to be a 32-byte $seed$ array much like that from Listing \ref{lst:safekeygen}. The secret key is then regenerated using the SHA-512 hash function, which is passed to the public key generation function, which performs the point multiplication. The rest of the Ed25519 signing function would then be implemented as per the standard.

\section{Conclusion}
\label{sec:conclusion}

In this work, an attack against the EdDSA standard is presented. Due to the deterministic nature of EdDSA signatures, an adversary with access to the signing function that accepts arbitrary public keys can recover the secret signing value by submitting as little as two different public keys. The adversary can sign arbitrary messages using this signing value, breaking the unforgeability security notion of digital signature schemes. The attack can occur primarily from software implementation APIs presenting an adversary's opportunity to submit multiple public keys to the signing function, creating different signatures for the same message and private key. This attack presents a real threat if applications expose these APIs publicly or fail to manage public-private key pairs correctly. A list of libraries that implement the Ed25519 standard and are vulnerable to this attack is given. Additionally, two countermeasures are proposed to prevent the attack.

\bibliographystyle{IEEEtran}

\bibliography{main}

\end{document}